\documentclass[11pt,twoside]{article}

\usepackage{tmpdps}
\usepackage{epsf}
\usepackage{psfig}
\usepackage{lscape}

\begin{document}
\title{Comparison of current models for Hot Jupiters to the sample of transiting exoplanets}  
\author{Michael Lund, Damian J. Christian}   

\affil{California State University, Department of Physics and Astronomy, Northridge, CA, USA, 91330}    


\begin{abstract} 
A growing number (over 100!) of extra-solar planets (ESPs)
have been discovered by transit photometry, and these systems are important because the transit strongly constrains their orbital inclination and allows accurate physical parameters for the planet to be derived, especially their radii.  Their mass-radius relation allows us to probe their internal structure.
In the present work we  calculate Safronov numbers for the current sample of ESP and compare their masses and radii to current models with the goal of obtaining better constrains on their formation processess. Our calculation of Safronov numbers for the current TESP sample does show 2 classes, although about 20\% lie above the formal Class I definition. These trends and recent results that argue against a useful distinction between Safronov classes are under further investigation. Mass-radius relations for the current sample of TESP are inconsistent with ESP models with very large core masses ($\geq$ 100 M$_{\oplus}$). Most TESP with radii near 1R$_J$ are consistent with models with no core mass or core masses of 10 M$_{\oplus}$. The inflated planets, with radii $\geq$1.2 R$_J$ are not consistent with current ESP models, but may lie along the lower end of models for brown dwarfs. Although such models are nascent, it is important to establish trends for the current sample of ESP, which will further the understanding of their formation and evolution. 
\end{abstract}

\section{Introduction}
After a slow start, photometric transit surveys have now provided over 100 transiting exoplanets (ESP) (as of October 2010). Tranist observations and follow-up radial velocity data strongly constrain the planets radii, masses and inclinations. Such accurate physical parameters then allow investigation of the internal structures of these planets and possible clues to their formation mechanisms.  The majority of transiting ESP (TESP) have masses below 1.5M$_J$, although there are several at masses near 3M$_J$ (e.g. WASP-10, HD 17156, and  CoRoT-2) and a similar number above 3M$_J$,
but  even with higher mass 
objects like the 7.3 M$_J$ WASP-14b \citep{Josh08} and 8 M$_J$ HD~147506b \citep{sato2005} this higher mass region has been poorly explored.  Most TESP had radii near 1R$_J$, but many  TESP had radii $\approx$ 10\% larger than
expected for their mass, like HD~209458b \citep{Br01,VM03}.

  There have been several attempts to explain these larger than expected radii.
 \citet{bar03} found irradiation effects could significantly alter the radii of sub-jovain mass giant planets, but not explain the large radii for the planets with radii greater than 1.2$R_J$, such as HD 209458b.  Additional authors have also considered
inflation due to irradiation effects  \citep{burrows2000, fortney2007} or other 
effects, like tidal forcing \citep{bode2003}, and internal dissipation of tidal energy arising
from orbital circularization \citep{bode2001}  but a consensus on an explanation for the inflated radii of these planets has not be reached. Regardless of these mass-radius discrepancies, photometric surveys have now provided a large sample of transiting ESP
that can be used to determine their mass-radius relation and test current models
for the structure and composition of ESP.

A recent parameterization to investigate possible classes of ESP has been presented by
 \citet{HB07} using the Safronov number for each system derived from the planet's equilibrium temperature, which is a function of its distance from its host star and the host stars effective temperature and radius. 
In this paper we present Safronov numbers for the current sample of transiting exoplanets (TESP) and a comparison of this sample to some of the latest models for mass-radus relationships for exoplanets. We selected the 100+ transiting extra-solar planets (TESP) listed in the Extrasolar planet encyclopedia\footnotemark ~as of October 2010. 

 \footnotetext{www.exoplanet.eu}

\begin{figure}[t]
\psfig{figure=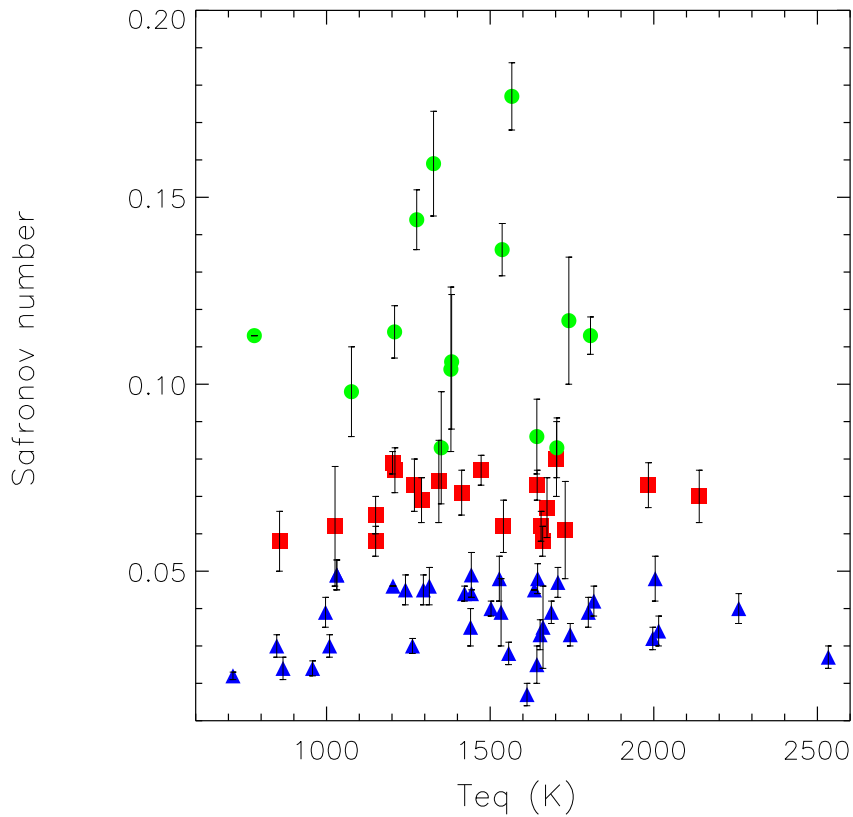,width=11.5cm,angle=0}
\caption[]{Plot of the Safronov number for the current  sample of transiting extra-solar planets  (as of April 2010) as a function of
each planet's equilibrium temperature. Class I planets with Safronov numbers in the $\approx$0.06 to 0.08 range are shown as red squares. Class II planets with Safronov numbers in the $\approx$0.03 to 0.05 range are shown as blue triangles. Lastly. ESP outside of these 2 ranges are shown as green circles (see text).  
}
\end{figure}

\begin{figure}
\vspace{-2.3in}
\begin{center}
\psfig{figure=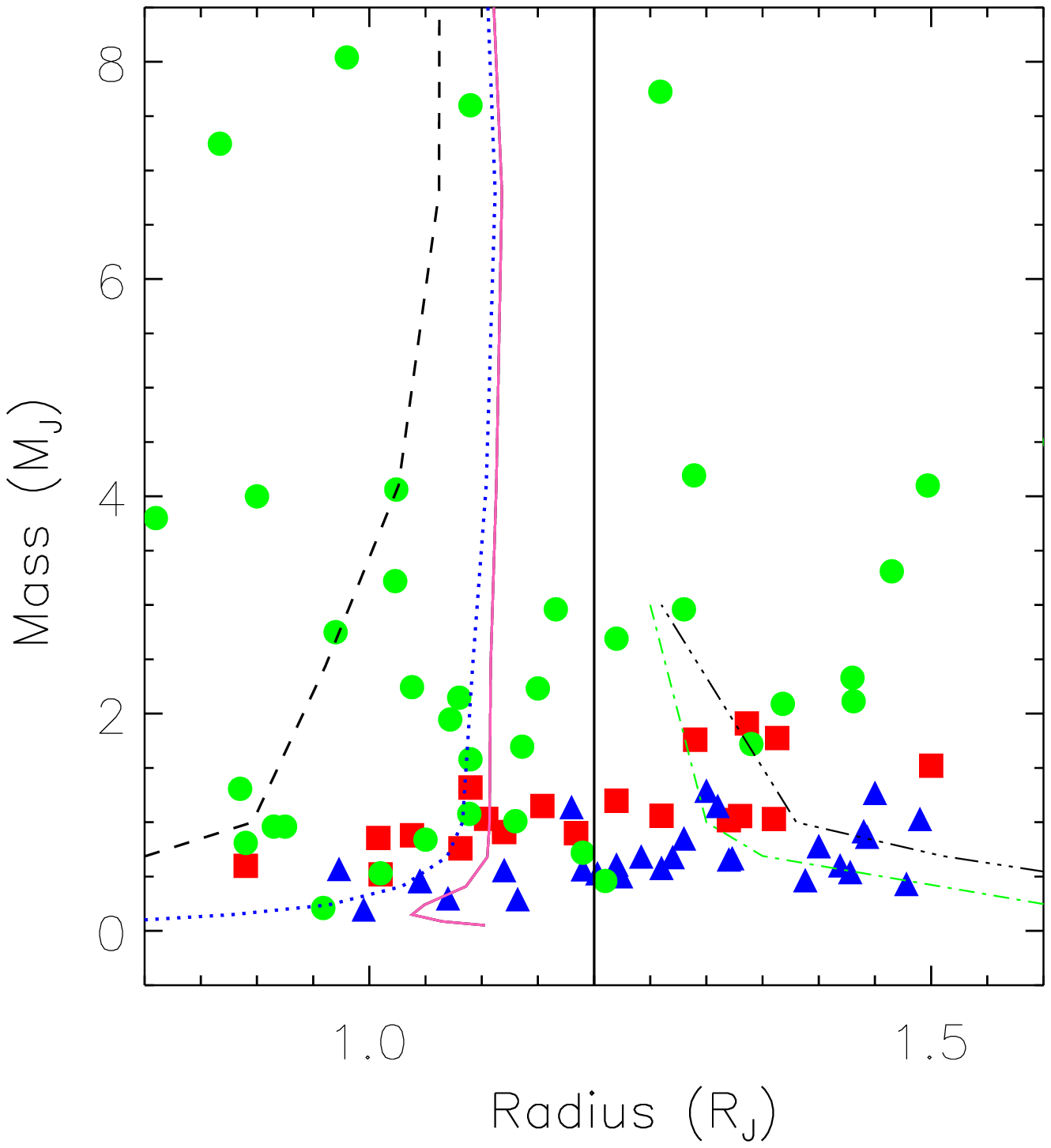,width=14.5cm,angle=0}
%
\vspace{-0.4in}
\caption[]{ Comparison of ESP models to the current sample of ESP.
Shown are the mass and radii 
with symbols based on Safronov number as defined in Figure 1. 
Over-plotted are the mass-radius relations from
\citet{fortney2007} using the 1.0 Gyr models and an orbital distance
of 0.045 AU. The core masses of 0, 10 and 100 M$_{\oplus}$, are shown as 
solid (pink), dotted (blue), and dashed (black) lines, respectively. 
The \citet{bode2003} models are shown on the right side of the figure, and are for a 4.5 Gyr planet with kinetic heating and an equilibrium temperature of 1500 K and for a planet with no core (long dashed black line), and with core (dashed green line).  
A vertical line at 1.2R$_J$ is also shown.
}
\label{fig:mrfortney}
\end{center}
\end{figure}

\section{Results \& Discussion}

We plot the Safronov number as a function of equilibrium temperature for the current sample of TESP in Figure 1.  Planets such as, HD~189733, TrES-3, HAT-P-7, and XO-4 can be seen in range of Safronov number between $\approx$0.06 to 0.08, noted as class I. ESP such as, GJ 436, WASP-11b, HD~209458b, and HD~149026b fall in Class II with Safronov number $\approx$0.03 to 0.05. There are a handful of plants have Safronov numbers higher than those of Class II shown in Figure 1 and these include WASP-7b and CoROT-Exo-2, and massive planets such as WASP-14b and HAT-P-2 (HD~147506b) are located off of the plot at Safronov numbers greater than 0.3. The figure clearly shows these 2 classes noted by \citet{HB07}. 

Our plot of the Safronov number as a function of  equilibrium temperature (Figure 1) does clearly show two classed of TESP. 
However, 
the significance of the gap observed in the Safronov numbers is still under scrutiny. \citet{So08} notes that since \citet{HB07}, additional extrasolar planet discoveries has led to the gap between the two classes of extrasolar planets weakening, and questions the statistical significance of the gap in Safronov numbers.  A more thorough evaluation of Safronov numbers, by \citet{Fres09}, contends that the gap is not presently statistically significant.  \citet{Fres09} also comment that there is no bimodal distribution of the two classes present in any other properties of extrasolar planets, raising questions as to if there actually are two separate classes of planets. More recently, Southworth 2010 claims there is no significance to Safronov numbers. The Kolmogorov-Smirnov test shows a low but significant deviation for Class I of 0.4 (K-S statistic) and Class II of 0.62 from the entire sample of 100 TESP planets, respectively. 

Comparison of current models for mass-radius relations \citep{burr97,bar03, bode2003, fortney2007} to the current sample of TESP (Figure 2) shows that the largest core masses have the smallest planetary radii and can be ruled out by the current sample of ESP.  A few ESP in the current sample are 
consistent with 10 M$_{\oplus}$ cores for the smallest objects
and several ESP with radii near $\approx$1 R$_J$ have masses consistent
with modes with no core mass.
However, planets with radii $>$ 1.2 R$_J$ are inconsistent with the
model behavior and may indicate another mechanism that 
inflates their radii \citep{mard07}. 
Curiously planets with radii in the $\approx$1.3--1.5 R$_J$ range do lie along the 
\citet{bode2003} models.  

The M-R plots hints there are two classes of ESP, those with moderate core masses and $\leq$1.2 R$_J$ and a second class of inflated ESP with radii $>$1.2 R$_J$ that are yet to be explained, although these may indicate another formation mechanism.  The usefulness of Safronov numbers is still being investigated.

\acknowledgements 
DC and ML thank the CSUN Department of Physics and Astronomy for support for this research. 



\begin{thebibliography}
\bibitem[Baraffe et al.(2008)]{bar08} Baraffe, I., Chabrier, G., Barman, T. 2008, A\&A, 482, 315, and arXiv:0802.1810  
\bibitem[Baraffe et al.(2003)]{bar03} Baraffe, I., Chabrier, G., Barman, T., Allard, N., \& Hauschildt, P.  2003, A\&A, 402, 701
\bibitem[Bodenheimer et al.(2001)]{bode2001}Bodenheimer, P., Lin, D.N.C., Mardling, R.A.  2001, ApJ, 548, 466
\bibitem[Bodenheimer et al.(2003)]{bode2003}Bodenheimer, P., Laughlin, G., Lin, D.N.C. 2003, ApJ, 592, 555
\bibitem[\protect\citeauthoryear{Brown et al.}{2001}]{Br01}Brown, T. et al. 2001, ApJ, 552, 699
\bibitem[Burrows et al.(2007)]{burrows2007} Burrows A., Hubeny I., Budaj J., Hubbard W.B., 2007, ApJ, 661, 502
\bibitem[Burrows et al.(2000)]{burrows2000} Burrows, A., Guillot, T., Hubbard, W.B., Marley, M.S., Saumon, D., Lunine, J.I., Sudarsky, D. 2000, ApJ, 34, L97
\bibitem[Burrows et al.(1997)]{burr97} Burrows A. et al. 1997, ApJ, 491, 856 
\bibitem[Fortney et al.(2007)]{fortney2007} Fortney J.J., Marley M.S., Barnes J.W. 2007, ApJ, 659, 1661
\bibitem[Fressin et al.(2009)]{Fres09} Fressin, F., Guillot, T., \& Nesta, L. 2009, A\&A,  504, 605
\bibitem[Hansen \& Barman(2007)]{HB07}Hansen, B.M., Barman, T. 2007, ApJ, 671 861 
\bibitem[\protect\citeauthoryear{Joshi et al.}{2008}]{Josh08}Joshi, Y.C. et al. 2008, MNRAS, 393, 1532
\bibitem[Mardling and references therein(2007)]{mard07}Mardling, R. 2007, MNRAS, 382, 1768
\bibitem[Sato et al.(2005)]{sato2005} Sato B., et al., 2005, ApJ, 633, 465
\bibitem[Southworth (2008)]{So08}  Southworth J. 2008, MNRAS, 386, 1644

\bibitem[Southworth (2010)]{So10}  Southworth J. 2010, MNRAS, in press.

\bibitem[Vidal-Madjar et al.(2003)]{VM03} Vidal-Madjar, A. et al. 2003, Nature, 422, 143 
\end{thebibliography}
\end{document}